\documentclass[twocolumn,showpacs,aps,amssymb,floatfix,prd,amsmath,preprintnumbers,10pt]{revtex4-2} 
\usepackage{graphicx}  
\usepackage{dcolumn}   
\usepackage{xcolor}
\usepackage{bm}
\begin{document}
\def \reals{{\mathbb R}}
\def\be{\begin{equation}}
\def\ee{\end{equation}}
\def\bea{\begin{eqnarray}}
\def\eea{\end{eqnarray}}
\def\nn{\nonumber}
\def\th{\theta}
\def\ph{\phi}
\def\lt{\left}
\def\rt{\right}
\def\degree{\mathop{\rm {{}^\circ}}}
\input epsf.tex
%%%%%%%%%%%%
\title{Conservation of distortion of gravitationally  lensed images}\thanks{This paper is  dedicated to my first  {\em guru} of general relativity
 Professor Ram Bahadur Singh of Patna University. (In Sanskrit, {\em gu} means darkness and {\em ru} means one who dispels.)}

\author{K. S. Virbhadra}
    \email[Email address : ]{shwetket@yahoo.com}
    \affiliation{Mathematics Department, Drexel University, 33rd and Market Streets, Philadelphia, Pennsylvania 19104, USA}
%%%%%%%%%%%%%%%%%%%%%%%%%%%%%%%%%%%%%%%%%%%%%%%%%%%%%%%%%%%%%%%%%%%%%%%%%%%%%%%%%%%%%%%%%%%%%%%%%%%%%%%%%%%%%%%%%%%%%%%%%%%
\begin{abstract}
We recently hypothesized that a distortion parameter exists such that its signed sum for all images of singular gravitational lensing of a source vanishes identically  [K. S. Virbhadra, Phys. Rev. D {\bf 106}, 064038 (2022)]. We found a distortion parameter (the ratio of the tangential to radial magnifications) that satisfied the hypothesis for the images of Schwarzschild lensing with flying colors. We now show that another distortion parameter (the difference of tangential and radial magnifications) also magnificently supports our hypothesis when we perform computations with the primary-secondary and relativistic images. The distortion parameters, which satisfy the aesthetically appealing hypothesis, will likely aid in developing gravitational lensing theory. Finally, we discuss the conservation of distortion of images in gravitational lensing.
\end{abstract}



\maketitle
%%%%%%%%%%%%%%%%%%%%%%%%%%%%%%%%%%%%%%%%%%%%%%%%%%%%%%%%%%%%%%%%%%%%%%%%%%%%%%%%%%%%%%%%%%%%%%%%%%%%%%%%%%%%%%%%%%%%%%%%%%%%%%%%%%%%%%%%%
\section{\label{sec:Intro}Introduction}
In 1959 (around two decades before the first gravitationally lensed images were observed), Sir Charles Darwin \cite{Dar58} pioneered gravitational lensing (GL) research due to the light deflection in the vicinity of a Schwarzschild {\em photohole} \cite{photohole}. He found that those images were too demagnified to be observed, and it seems that he referred to them as ``ghosts" because of this. Research on GL  due to light deflection in a very strong gravitational field remained almost stagnant for nearly four decades until we\cite{VE00} obtained a new gravitational lens equation. 
This equation enables the study of light deflection in weak and strong gravitational fields, including those near photon spheres.
Unaware of Darwin's work, we studied GL due to light deflection near the photon sphere of Schwarzschild ``black hole". Like Darwin, we also found images that were very demagnified and termed them ``relativistic images" (lensed images due to light deflection $\hat{\alpha} > 3 \pi/2$). 
Despite the discouraging theoretical results of very demagnified relativistic images, our work revived theoretical research on gravitational lensing by massive compact objects, such as black holes and other compact esoteric objects (see in \cite{CVE01,Vir09,YHJB15,ISOKA16,BT17,Tsu17,ISOA17,OIA17,ZX17,VE02,VK08,GY08,NIA14,BRS14,JBGA19}  and references therein.)
 The monumental success of the  Event Horizon Telescope ({\em EHT}) project in 2019  (surprisingly around one hundred years after the first light deflection was observed under the leadership of Eddington) \cite{EHT6} as well as the ongoing development of the  Next Generation Event Horizon Telescope {\em ngEHT} \cite{ngEHT,JohnEtal23}  project have generated significant theoretical interest in  GL due to black holes and  their mimickers  (see \cite{TOA20,KIG20,HLL21,Tsu21,CORR21,ABN21, AJJAA21,AAH21,GCZYH21,CJL21,KBG22,KIG22,Tsu22,AV22,AJM22,BGGKNV22,BP22,GB22,GJWW22,POD23,Tsu23,Wal23,PPO23,BKG23,PK23,SPVR23,RPO23,MMDH23,KO23,OPR23,NS23,LZLLS23,RIPA23,GYH23,NQ23,BSG23,KBG24,JLDMXZW24,SR20,GKNTY20,GX21,NCC23,GLM20,LCCH21}
and references therein.)

The theory of GL  and its implications for astrophysics and cosmology are well discussed in \cite{Book,NB96}. We \cite{CVE01}  proved important theorems on photon surfaces that have significant implications for GL. In our recent paper\cite{Vir22}, we hypothesized that a distortion parameter exists such that its signed sum for all images of a singular GL of a source identically vanishes. We provided a distortion parameter and demonstrated that our hypothesis is valid. Toward the end of the paper, we proposed another distortion parameter. We now show that the second distortion parameter also satisfies the distortion hypothesis with all images, including relativistic images, and this is the plan for this paper. Similar to our previous paper on the distortion hypothesis  \cite{Vir22}, we use the geometrized units so that the ADM (Arnowitt-Deser-Misner) mass parameter $M \equiv MG/c^2$, where $G$ is the universal gravitational constant and $c$ is the speed of light in vacuum. We use {\it Mathematica} \cite{Math13} for computations.

%%%%%%%%%%%%%%%%%%%%%%%%%%%%%%%%%%%%%%%%%%%%%%%%%%%%%%%%%%%%%%%%%
\section{\label{sec:LE-DP}Lens Equation and Distortion Hypothesis}

A gravitational lens equation that allows arbitrarily large as well as small deflection angles of light rays is given by \cite{VE00}

%%%%%%%%
\be
\tan\beta = \tan\theta - \alpha ,
\label{LensEqn}
\ee
%%%%%%
where
\be
\alpha =
{\cal D} \lt[\tan\theta + \tan\lt(\hat{\alpha} - \theta\rt)\rt]
\label{Alpha}
\ee
%%%%%%%
with a dimensionless {\em distance parameter}
%%%%%%
\be
{\cal D} = \frac{D_{ds}}{D_s} \in \left(0,1\right) \text{.}
\ee
%%%%%%%%%%%%
The symbols have the same meaning as in our previous paper\cite{Vir22}:  $D_s$, $D_{ds}$, and $D_d$, represent, respectively, the observer-source, deflector (lens)-source, and observer-deflector angular diameter distances, whereas $\beta$ and $\theta$ stand for the angular source and image positions, respectively. For $0.5<{\cal D}<1$, the source is farther (from the lens) compared to the observer, and for $0<{\cal D}<0.5$, it is nearer. For ${\cal D} = 0.5$, the lens is halfway between the observer and the source.

The total magnification of an image of circularly symmetric gravitational lensing, which can have positive or negative values, is given by  \cite{Book}
%%%%%%%%
\be
\mu = \mu_r \mu_t \text{,}
\label{Mu}
\ee
%%%%%%%%%
where the radial $\mu_r$ and tangential $\mu_t$ magnifications are, respectively, given by
\be
\mu_r = \lt(\frac{d\beta}{d\theta}\rt)^{-1} ~ ~ ~ \text{and} ~ ~ ~
\mu_t = \lt(\frac{\sin{\beta}}{\sin{\theta}}\rt)^{-1} \text{.}
\label{MutMur}
\ee
The following line element describes the exterior gravitational field of a Schwarzschild black hole:
%%%%%%%%%%%%%%%%%%%%%%
\begin{eqnarray}
ds^2&=&\lt(1-\frac{2M}{r}\rt)dt^2- \lt(1-\frac{2M}{r}\rt)^{-1} dr^2 \nonumber \\
&-&r^2\lt(d\vartheta^2+\sin^2 \vartheta d\phi^2\rt) \text{.}
\label{SchMetric}
\end{eqnarray}
The constant parameter $M \equiv MG/c^2$ is the ADM mass of the black hole.
Defining the scaled radial distance and the scaled closest distance of approach, respectively, by
\begin{equation}
\rho = \frac{r}{2M} , ~ ~ ~ \text{and} ~ ~ ~
\rho_o = \frac{r_o}{2M}
\label{XX0}
\end{equation}
%%%%%%%
($r_0$ is the closest distance of approach), the Einstein bending angle $\hat{\alpha}$ and the impact parameter $J$ of a light ray\cite{Wei72,VNC98,VE00}:
\begin{equation}
\hat{\alpha}\lt(\rho_o\rt) = 2 \ {\int_{\rho_o}}^{\infty}
\frac{d\rho}{\rho \ \sqrt{\lt(\frac{\rho}{\rho_o}\rt)^2
\lt(1-\frac{1}{\rho_o}\rt)
-\lt(1-\frac{1}{\rho}\rt)}} - \pi
\label{AlphaHatRho0}
\end{equation}
%%%%%%%%%%
and
%%%%%%%
\begin{equation}
J\lt(\rho_o\rt) = 2M \rho_o \lt(1-\frac{1}{\rho_o}\rt)^{-1/2}.
\label{ImpParaRho0}
\end{equation}
Around two years ago, we \cite{Vir22} hypothesized that a distortion parameter exists such that its signed sum for all images of a singular GL of a source identically vanishes. To demonstrate the correctness of this hypothesis,
we introduced a {\em distortion parameter} for a gravitationally lensed image:
\be
\boxed{\Delta = \frac{\mu_t}{\mu_r} }
\label{Delta}
\ee
%%%%%%%%%%%%%%%
and the signed sum of distortions of all images of a given source
%%%%%%%%%%%%
\be
{\Delta}_{sum} = \sum_{i=1}^{k} \Delta_i \text{,}
\label{SumDistortions}
\ee
%%%%%%%%%%%%%%
where $k$, the upper limit of the summation, represents the total number of images. (Hereafter, we will call $\Delta$ the {\em first distortion parameter}.) We carried out computations for primary-secondary as well as relativistic images and showed that the hypothesis is correct. We mentioned that there could be more than one such distortion parameter and we \cite{Vir22} proposed another one (henceforth, we will call it the {\em second distortion parameter} ) as follows:
\be
\boxed{\overset{\star}{\Delta}= \mu_t-\mu_r \text{.}}
\label{Delta}
\ee

% \be
% \boxed{\Delta^{\star}= \mu_t-\mu_r \text{.}}
% \label{Delta}
%\ee
Then we computed this quantity for the primary as well as the secondary images under weak-field approximation:
\be
\overset{\star}{\Delta}_p = - \overset{\star}{\Delta}_s = \frac{8 {\cal D} \frac{M}{D_d} } {\beta \sqrt{16 {\cal D} \frac{M}{D_d} + \beta^2}} \text{.}
\ee
(The subscript $p$ and $s$, respectively, stand for the primary and secondary images.) Thus, the signed sum of these distortions
\be
\overset{\star}{\Delta}_{sum} = \sum_{i=1}^{k} \overset{\star}{\Delta}_i = \overset{\star}{\Delta}_p + \overset{\star}{\Delta}_s = 0 \text{,}
\label{SumDistortions}
\ee
where $k=1,2$, $\overset{\star}{\Delta}_1 = \overset{\star}{\Delta}_p$, and $\overset{\star}{\Delta}_2=\overset{\star}{\Delta}_s$. (Note that in our previous paper\cite{Vir22}, we did not put $\star$ as overscripts.) The subscripts $p$ and $s$, respectively, stand for the primary (also called the direct) and secondary images. Though this is encouraging, it is not enough to support the distortion hypothesis. Therefore, we perform numerical computations without weak- or strong-field approximation for the primary-secondary and relativistic images. We do this in the next section. To conveniently compare distortions of images of several orders, we define
a {\em logarithmic distortion parameter} of an image
%%%%%%%%%%%%
%%%%%%%%%%%%%%%%%%%%
\be
\overset{\star}{\delta} = \log_{10} \big| \overset{\star}{\Delta} \big| \text{.}
\label{delta}
\ee
%%%%%%%%%
The images of the same order are on opposite sides of the optic axis and their distortions have opposite signs. To test whether images of the same order have the same absolute distortions, we now define {\em percentage difference in distortions} of images of the same order as follows:
%%%%%%%%%%%%%%%%
\bea
\overset{\star} {\mathbb{P}}_{ps} &=& \frac{|\overset{\star}{\Delta}_p| -|\overset{\star}{\Delta}_s|}{|\overset{\star}{\Delta}_p|}\times 100 \text{,} \nn \\
\overset{\star} {\mathbb{P}}_{1p1s} &=& \frac{|\overset{\star}{\Delta}_{1p}| -|\overset{\star}{\Delta}_{1s}|}{|\overset{\star}{\Delta}_{1p}|}\times 100 \text{,} \nn \\
%\frac{|\Delta^{\star}_{1p}|-|\Delta^{\star}_{1s}|}{|\Delta^{\star}_{1p}|}\times 100 \text{,} \nn \\
\overset{\star} {\mathbb{P}}_{2p2s} &=& \frac{|\overset{\star}{\Delta}_{2p}| -|\overset{\star}{\Delta}_{2s}|}{|\overset{\star}{\Delta}_{2p}|}\times 100 \text{,}
% \frac{|\Delta^{\star}_{2p}| -|\Delta^{\star}_{2s}|}{|\Delta^{\star}_{2p}|}\times 100 \text{,}
\label{PercentageDiff}
\eea
%%%%%%%%%%%%%%%%%%
The subscripts $ps$, $1p1s$, and $2p2s$, respectively, stand for the primary-secondary, relativistic image pairs of the first order and of the second order. (The subscripts $1$ and  $2$ represent relativistic images of the first and second orders, respectively.)
%%%%%%%%%%%%%%%%%%%%%%%%%%%%%%%%%%%%%%%%%%%%%%%%%%%%%%%%%%%%%%%%%%%%
% %%%%%%%%%%     Figure 1     %%%%%%%%%%%%%%%%%%%%%%%%%%%%%
%%%%%%%%%%%%%%%%%%%%%%%%%%%%%%%%%%%%%%%%%%%%%%
\begin{figure*}[tbh]
\centerline{
     \epsfxsize 6.20cm  \epsfbox{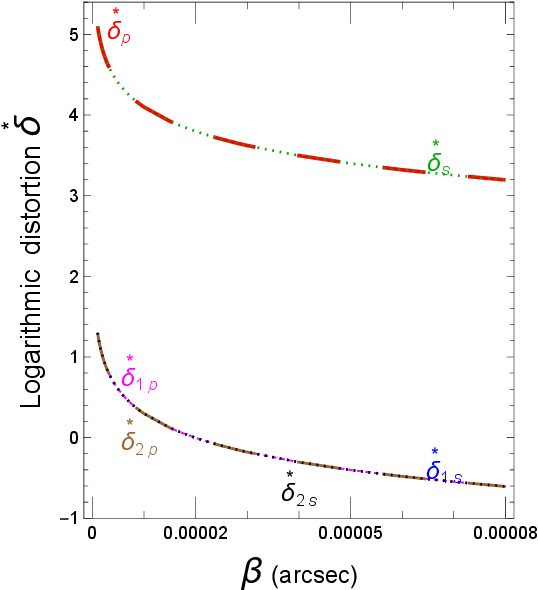}  \qquad      \epsfxsize 5.92cm  \epsfbox{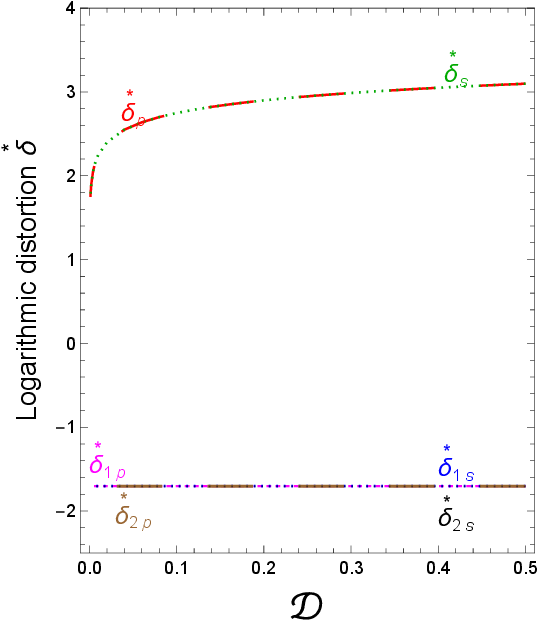} \qquad      \epsfxsize 5.92cm  \epsfbox{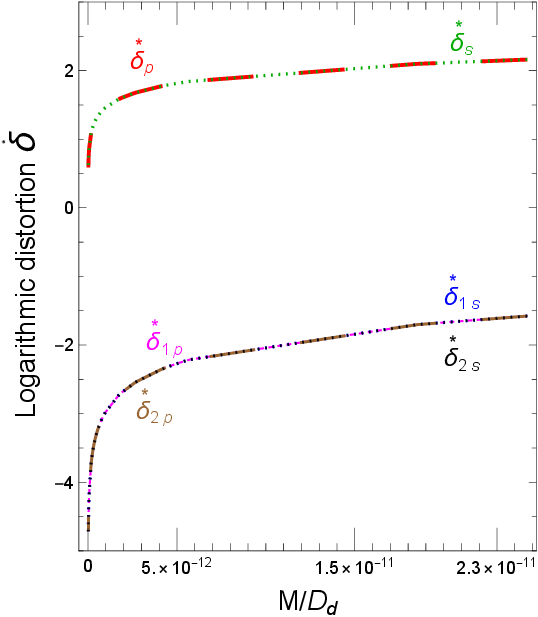}}
 \caption[ ]{
{\em Left }: the Schwarzschild lens model describes the supermassive dark object at the galactic center of M87 with the distance parameter ${\cal D} = 0.005$. The lens has mass $M \approx 6.5 \times 10^9 M_{\odot}$ and is at the distance $D_d \approx  16.8$ {\em Mpc}.  The logarithmic distortions (of the second type) of the primary image $\overset{\star}{\delta_p}$ (red dashed), secondary image $\overset{\star}{\delta_s}$ (green dotted), the first-order relativistic image on the primary side $\overset{\star}{\delta_{1p}}$ (magenta dashed), the first-order relativistic image on the secondary side $\overset{\star}{\delta_{1s}}$ (blue dotted), the second-order relativistic image on the primary side $\overset{\star}{\delta_{2p}}$ (brown dashed), and the second-order relativistic image on secondary side $\overset{\star}{\delta_{2s}}$ (black dotted) are plotted against the angular source position $\beta$. 
%%%%%%%%%%%%
{\em Middle}:   for the same lens but the angular source position $\beta = 1 mas$, the exact six quantities (as for the figure on the left) are plotted against the parameter ${\cal D} $. 
%%%%%%%%
{\em Right}:  the galactic centers of $40$ supermassive dark objects are modeled as    Schwarzschild lenses. For the distance parameter $ {\cal D} = 0.005$ and the angular source position $ \beta = 1 mas $, the exact six quantities (as for the figures on the left and the middle) are plotted against the potential $M/D_d$. The colors of symbols and graphs are kept the same to identify graphs.}
\label{fig1}
\end{figure*}
%%%%%%%%%%%%%%%%%%%%%%%%%%%%%%%%%%%%%%%%%%%%%%
%%%%%%%%%%%%  Figure 2   %%%%%%%%%%%%%%%%%%%%%%%%%%%%%
%%%%%%%%%%%%%%%%%%%%%%%%%%%%%%%%%%%%%%%%%%%%%%%
\begin{figure*}[tbh]
\centerline{
     \epsfxsize 6.0cm  \epsfbox{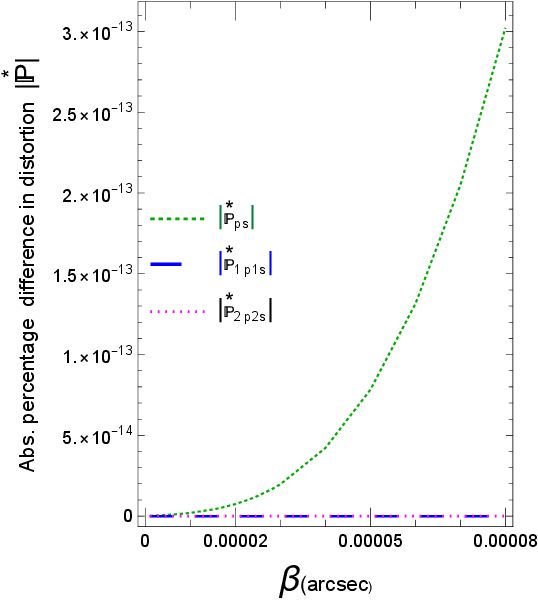}  \qquad      \epsfxsize 5.5cm  \epsfbox{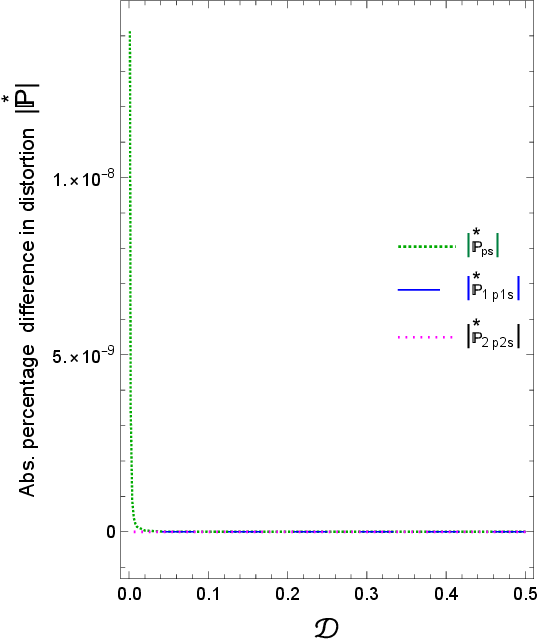} \qquad      \epsfxsize 5.5cm  \epsfbox{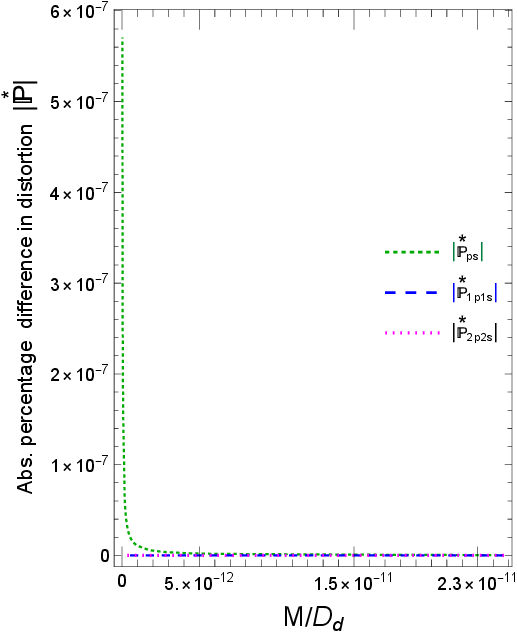}
          }
 \caption[ ]{
{\em Left}:  
the supermassive dark object at the center of M87 is modeled as the Schwarzschild lens, which has mass $M \approx  6.5 \times 10^9 M_{\odot}$ and is at a distance $D_d \approx  16.8$ {\em Mpc}.  The dimensionless parameter ${\cal D} = 0.005$ gives the source position. The absolute percentage difference in distortions (of the second type) for the primary-secondary images pair $\lvert\overset{\star}{\mathbb{P}}_{ps}\rvert$, the first-order relativistic images pair  $\lvert \overset{\star}{\mathbb{P}}_{1p1s}\rvert$,  and the second-order relativistic images pair  $\lvert\overset{\star}{\mathbb{P}}_{2p2s}\rvert$ vs the angular source position $\beta$ are plotted.
%%%%%%%%%%%%
{\em Middle}: the exact three quantities (as for the figure on the left)  against the parameter ${\cal D}$ are plotted. The lens is the same; however,  the angular source position $\beta = 1 mas$.  
%%%%%%%%
{\em Right}: the exact three quantities are plotted against the potential $M/D_d$ as for the figure in the left and in the middle. The supermassive dark objects at the centers of 40 galaxies are modeled as  Schwarzschild lenses with the angular source position $\beta = 1 mas$ and the distance parameter  ${\cal D} = 0.005$.
 }
\label{fig2}
\end{figure*}
%%%%%%%%%%%%%%%%%%%%%%%%%%%%%%%%%%%%%%%%%%%%%
\section{\label{sec:Comp-Res}Computations and Results}
In \cite{Vir22}, we modeled $M87^{\star}$ (the potential $M/D_d \approx 1.84951 \times 10^{-11}$) as the Schwarzschild lens with dimensionless distance parameter ${\cal{D}} = 0.005$. For several values of the angular source position $\beta$, we performed computations for the tangential and radial magnifications of the primary-secondary and relativistic images of orders $1$ and $2$. We repeated these computations with the fixed angular source position $\beta=1 mas$ and $M/D_d = 1.84951 \times 10^{-11}$ for a large number of values for the distance parameter ${\cal{D}}$. Finally, we modeled the galactic centers of $40$ galaxies as the Schwarzschild lenses and repeated the computations with $\beta= 1 mas$ and ${\cal{D}} = 0.005$. Now, using these results for the tangential and radial magnifications ($\mu_t$ and $\mu_r$), we compute the second logarithmic distortion $\overset{\star} {\delta}$ for  images. In Fig. 1, we plot the logarithmic distortions $\overset{\star} {\delta}$ against $\beta$ ( ${\cal{D}} = 0.005$ and $M/D_d \approx 1.84951 \times 10^{-11}$ fixed), $\overset{\star} {\delta}$ against ${\cal{D}}$ ( $\beta=1 mas$ and $M/D_d \approx 1.84951 \times 10^{-11}$ fixed), and $\overset{\star} {\delta}$ against the potential $M/D_d$ ($\beta = 1 mas$ and ${\cal{D}} = 0.005$ fixed) for the set of six images (primary-secondary as well as relativistic images of order $1$ and $2$ on both sides of the optic axis). These plots show that the logarithmic distortions $\overset{\star} {\delta}$ of images of the same order are too close to be resolved, supporting the distortion hypothesis. The figures also show that logarithmic distortions $\overset{\star} {\delta}$ are higher for the lower order of images. (This is contrary to the case logarithmic distortion $ {\delta}$ of the first type.) For example, the primary-secondary pair (images of order zero) $\overset{\star} {\delta}$ is the highest. The $\overset{\star} {\delta}$ vs $\beta$ graph shows that $\overset{\star} {\delta}$  first decreases fast and then slowly with the increase in the value of the angular source position $\beta$ and curves are concave up. This is the same as the case for the first logarithmic distortion $ {\delta}$). $\overset{\star} {\delta}$ vs ${\cal{D}}$ graphs show that $\overset{\star} {\delta}$ increases fast for small values of ${\cal{D}}$ and then slowly, and all curves are concave down. This is similar to the case of the first logarithmic distortion variations. The $\overset{\star} {\delta}$ vs the potential $M/D_d$ curves increase fast and then slowly; all curves are concave down. These results are again similar to the case of the first logarithmic distortion.

Figure 1 shows that the magnitudes of logarithmic distortions of images of the same order are very close. As $\overset{\star} {\Delta} = \mu_t-\mu_r$ for images of the same order have opposite signs, this strongly supports the distortion hypothesis. Now, we would like to know the percentage difference in magnitudes of distortions to explore whether these are due to assumptions involved in the lens equation and approximations  in numerical methods. Therefore, we now compute $\overset{\star} {\mathbb{P}}_{ps}$, $\overset{\star} {\mathbb{P}}_{1p1s}$, and $\overset{\star} {\mathbb{P}}_{2p2s}$ given in  Eq.  $(\ref{PercentageDiff})$  for the same six images. In Fig. 2, we then plot $\overset{\star} {\mathbb{P}}_{ps}$ vs $\beta$ (keeping ${\cal{D}} = 0.005$ and $M/D_d \approx 1.84951 \times 10^{-11}$ fixed), $\overset{\star} {\mathbb{P}}_{1p1s}$ vs ${\cal{D}}$ (keeping $\beta=1 mas$ and $M/D_d \approx 1.84951 \times 10^{-11}$ fixed), and $\overset{\star} {\mathbb{P}}_{2p2s}$ vs $M/D_d$ (keeping $\beta=1 mas$ and ${\cal{D}} = 0.005$ fixed) for the same six images. As the percentage differences are too small,  these seem to be due to lens approximation and approximations with numerical methods. We argue that these should be zero if we used an exact lens equation and exact analytical methods were available. Thus, our second distortion parameter also supports the distortion hypothesis with flying colors.  Henceforth, we will refer to any function of radial and tangential magnifications as a distortion parameter if it satisfies our hypothesis.

%%%%%%%%%%%%%%%%%%%%%%%%%%%%%%%%%%%%%%%%%%%%%%%%%%%%%%
\section{\label{sec:Disc-Sum} Discussion and Summary}
In theoretical physics, in addition to matching experimental/observational facts and predicting results, we also consider the logical beauty of the laws when discovering laws of physics. In view of this, our {\em distortion hypothesis} originated as follows. We assigned distortion of an unlensed source zero. [The second distortion parameter $\overset{\star} {\delta}$  is clearly zero for an unlensed source. However,  we need to  slightly modify the  first distortion parameter as  ${\delta} = (-1)^l \log_{10}\left(\mu_t/\mu_r\right)$, where $l$ is the parity of image.]
 When a source is lensed by a spacetime, it may give rise to multiple images. Because the unlensed source has zero distortion, we came up with the idea that the signed sum of the distortion of all images remains zero. Based on this thought, we hypothesized: There exists a distortion parameter such that the signed sum of all images of a singular gravitational lensing of a source identically vanishes \cite{Vir22}. Fortunately, we found a distortion parameter $\Delta ={\mu_t } /{\mu_r}$ that satisfied the distortion hypothesis with the primary-secondary images under the weak-field approximation. This result encouraged us to conduct computations, including relativistic images without weak or strong gravitational field approximations, and our hypothesis won with flying colors. However, we were still determining if only one distortion parameter could exist. We attempted and discovered a second distortion parameter $\overset{\star} {\Delta} = \mu_t-\mu_r$ (the difference of the tangential and radial magnifications) and noticed that the signed sum for the primary-secondary images under the weak-field approximation identically vanishes. However, we needed to determine whether the hypothesis holds well if we carried out computations without weak- or strong-filed approximations, including relativistic images. Our second distortion parameter also succeeds with flying colors, so such a distortion parameter is not unique.
These aesthetically appealing results are likely to have significant implications for astronomy and astrophysics, especially in the development of gravitational lensing theory. We usually identify images of the same source by similar spectra, knots, and the same radio and optical wave-band flux ratio. However, as pointed out in our previous paper, there is no way to know if there are any missing images, and our distortion hypothesis might help locate those. We found that distortions of images of the same order on the opposite sides of the optic axis have equal magnitudes with opposite signs. These results suggest that images of the same order on the opposite sides of the optic axis might have quite different magnifications but have ``similar" shapes with opposite parities. The geometrical implications for both distortion parameters are worth studying.
The hypothesis should also be tested with the Kerr-Newman metric. Any linear combination of distortion parameters is obviously a distortion parameter. It is still to be determined whether there can be more than two linearly independent distortion parameters that we have discovered.  
As we assign the distortion of an unlensed source to be zero,  the signed sum of distortions of the source plus images is zero. Let us call this {\em conservation of distortion of images}.   To analytically prove or disprove our distortion hypothesis, we need to develop advanced-level mathematics, which is worth doing to improve the gravitational lensing theory. We used the term ``singular" gravitational lensing in our distortion hypothesis, which means that our hypothesis applies to singular GL, and may not be true for nonsingular lensing. If the conservation of distortion of images is indeed a physical law, then we need to find a distortion parameter that works for all gravitational lenses.

\acknowledgments
I  thank the anonymous referees for carefully reading the manuscript.
%%%%%%%%%%%%%%%%%%%%%%%%%%%%%%%%%%%%%%%%%%%%%%%%%%%%%%%%%%%%%%%%%%%%%%%%%%%%%%%%%%%%%%%%%%%%%%%%%%%%%%%%%%%%%%%%%%%%%%%%%%%%%%%%%%%%%%%%%%%%%%%%%%%%%

\end{document}